\documentclass[10pt]{llncs}       %--- LATEX 2e base
\usepackage{graphicx,subfigure}
\usepackage{epsfig}
\usepackage{algorithm}
\usepackage[noend]{algorithmic}
\usepackage{setspace}
\usepackage{amsfonts}
\usepackage{amsmath}

\newcommand{\keywords}[1]{\par\addvspace\baselineskip
\noindent\keywordname\enspace\ignorespaces#1}

\begin{document}

%% \twocolumn[
%%     \begin{@twocolumnfalse}

\title{NAPX: A Polynomial Time Approximation Scheme for the Noah's Ark Problem}
\author{Glenn Hickey\inst{1}
 \and Paz Carmi\inst{1} \and Anil Maheshwari\inst{1}
 \and Norbert Zeh\inst{2}}
\institute{School of Computer Science, Carleton University, 
 Ottawa, Ontario, Canada \and
Faculty of Computer Science, Dalhousie University, Halifax, Nova Scotia, 
Canada\\
\email{ghickey@scs.carleton.ca, carmip@gmail.com, anil@scs.carleton.ca,
nzeh@cs.dal.ca}}
\maketitle

\begin{abstract}
The Noah's Ark Problem (NAP) is an NP-Hard 
optimization problem with relevance to
ecological conservation management. It asks to maximize the phylogenetic
diversity (PD) of a set of taxa given a fixed budget, where each taxon
is associated with a cost of conservation and a probability of extinction.
NAP has received renewed interest with
the rise in  availability of genetic sequence data, allowing
PD to be used as a practical measure of biodiversity.  However, only
simplified instances of the problem, where one or more parameters are 
fixed as constants, have as of yet 
been addressed in the literature.  We present NAPX, the first algorithm
for the general version of NAP that returns a $1 - \epsilon$ approximation
of the optimal solution.  It runs in 
$O\left(\frac{n B^2 h^2 \left(\log{n} + \log{\frac{1}{\epsilon}}\right)^2}
{\log^2(1 - \epsilon)}\right)$ 
time where $n$ is the number of species, 
and $B$ is the total budget and $h$ is the height of the input tree. 
We also provide improved bounds for its expected running time.
\end{abstract}

\section{Introduction}
\subsection{Motivation}

Measures of biodiversity are commonly used as indicators of environmental 
health.  Biodiversity is presently being lost at an alarming rate, 
due largely to human activity.  It is speculated that this loss can lead
to disastrous consequences if left unchecked \cite{nee97}.  
Consequently, the discipline of conservation biology has arisen and
a considerable amount of resources are being allocated to research
and implement conservation projects around the world.

A conservation strategy will necessarily 
depend on the measure of biodiversity used. 
Traditionally, indices based on species richness and abundance have been
used to quantify the biodiversity of an ecosystem \cite{magurran04}. 
These indices are based on counting and do not account for genetic variance.
Phylogenetic diversity (PD) 
\cite{faith92} addresses this issue by taking into account evolutionary
relationships derived from DNA or protein samples.  
The use of PD in biological conservation has become 
increasingly widespread as more phylogenetic information becomes 
available \cite{heard00}.  It is also used to determine diverse
sequence samples in comparative genomics \cite{pardi05}. 

The Noah's Ark Problem (NAP) \cite{weitzman98} is an abstraction of the
fundamental problem of many conservation projects:  how best to allocate
a limited amount of resources to maximally conserve phylogenetic diversity.  
This
is in turn a generalization of the Knapsack Problem \cite{garey79} and
is therefore NP-Hard.  Several algorithms have been proposed to solve special
cases of the problem but, as yet, no non-heuristic solutions have been
proposed to solve general instances of NAP.  Given that NAP itself is
an abstraction of realistic scenarios, it is important to have a general
solution in order to be able to extend this framework
 to useful applications.  For
this reason, we present an algorithm that
can be used to compute an approximate solution for NAP
in polynomial time, so long as the approximation factor is held constant, 
and total budget is polynomial in the input size.

\subsection{Definitions}

Throughout this paper, we use the following definition
of a phylogenetic tree $\mathcal{T}$, with notation consistent
with that of \cite{hartmann06}.  $\mathcal{T}$ has a root of 
degree 2, interior vertices of degree 3, and $n$ leaves, each associated
with a species from set $X$. If an edge $e$ of $\mathcal{T}$
is incident to a leaf, it is
called a pendant edge.  Otherwise $e$ has exactly two adjacent edges, 
$l$ and $r$, below it (not on the path from $e$ to the
root) and these are referred to as $e$'s children. 
$\lambda$ is a function that assigns a  non-negative branch length to each
edge in $\mathcal{T}$.  The phylogenetic
diversity of $\mathcal{T}$, PD($\mathcal{T}$) is defined as
\begin{equation}
\label{eq:pd}
PD(\mathcal{T}) = \sum_e \lambda(e),
\end{equation}
where the summation is over each edge $e$ of the tree.  Intuitively, this
measure corresponds to the amount of evolutionary history represented 
by $\mathcal{T}$.

The Noah's Ark Problem has the
objective of 
maximizing the expected PD, $\mathbb{E}(PD)$, under the following
constraints.  Each taxon $i \in X$ is associated with an initial survival 
probability $a_i$, which can be increased to $b_i$ at some integer cost $c_i$;
and the total expenditure cannot exceed the budget $B$.  Since $B$ is a 
factor in the running time, we assume that the budget and each cost have been
divided by the greatest common divisor of all the costs. 
In the original
formulation of NAP, each species was also associated with a utility
value. However, 
in \cite{hartmann06} it was shown that
these values are redundant as they can
be incorporated into the branch lengths without altering the problem.
To avoid accounting for degenerately small probability values, we make the
assumption that the conserved survival probabilities are not exponentially
 small in $n$. In
other words, there exists a constant $k$ such that $b_i \geq n^{-k}$ for each
$i \in X$.
We feel this assumption is reasonable as it is unrealistic that money would
be allocated to obtain such a negligible probability of survival.

If a species survives, the information represented by its path to the root
is conserved.  Consequently, the probability that an edge survives is 
equivalent to the probability that at least one leaf below it in 
$\mathcal{T}$ survives.  Let $C_e$ be the set of leaves below $e$ in the
tree and $S$ be the set of species selected for 
protection.  
$\mathbb{E}(PD|S)$, can be derived from 
(\ref{eq:pd})  as follows: 
\begin{equation}
\label{eq:epd}
\mathbb{E}(PD|S) = \sum_e \lambda(e) \left(1 - \prod_{i \in C_e \cap S}
\left(1 - b_i \right) \prod_{j \in C_e - S}
\left(1 - a_j \right) \right),
\end{equation}
where the summation is over all edges. NAP asks to maximize $\mathbb{E}(PD|S)$
subject to
\begin{equation*}
\label{eq:bugetconstraint}
\sum_{s \in S}c_s   \leq B.
\end{equation*}
Our algorithm is based on decomposing $\mathcal{T}$ into clades which are
associated with the edges of the tree. 
A clade corresponding to edge $e$, denoted $\mathcal{K}_e$,
 is the minimal subtree of $\mathcal{T}$
containing $e$ and $C_e$, the set of leaves below it.  The $\mathbb{E}(PD)$ of 
$\mathcal{K}_e$ can be computed
as in (\ref{eq:epd}) but summing only over edges in the clade.  
The entire tree can be considered a clade by
attaching an edge of length 0 to its root.  If $e$ has two 
descendant edges $l$ and $r$,
then we say $\mathcal{K}_e$ has two 
child clades $\mathcal{K}_l$ and $\mathcal{K}_r$. 

\subsection{Related Work}

Let $a_i \xrightarrow{c_i} b_i$ NAP refer to the problem as described above,
where the survival probabilities and cost of each taxon are input variables.
Fixing one or more of these variables as constants 
produces a hierarchy of increasingly
simpler subproblems \cite{pardi07}.  The simplest, $0 \xrightarrow{1} 1$ NAP, 
is equivalent to finding the set of $B$ leaves whose induced subtree 
(including the root) has maximum PD and can be solved by a greedy
algorithm \cite{steel05} \cite{pardi05}. $0 \xrightarrow{c_i} 1$ NAP on
ultrametric (all leaves equidistant from the root)
 trees and $(1-x_i) \xrightarrow{1} (1 - \kappa x_i)$ for general
trees where $x_i$ is a variable probability and $\kappa$ is a 
constant factor such that $0 \leq \kappa \leq 1$ can likewise be 
solved in polynomial time by greedy algorithms~\cite{hartmann06}. 
Given that $0 \xrightarrow{c_i} 1$ NAP is itself a generalization
of the Knapsack problem which is NP-Hard, it is extremely unlikely that
an exact, polynomial-time solution for this kind of NAP or any generalizations
will ever be found.
Pardi and Goldman 
\cite{pardi07} did find a pseudopolynomial-time dynamic programming
algorithm for 
the $0 \xrightarrow{c_i} 1$ NAP on general (non-ultrametric) trees
that makes the realistic assumption that $B$ is polynomial in $n$.
They
also show that any instance of $a_i \xrightarrow{c_i} 1$ NAP can be
transformed to an instance of $0 \xrightarrow{c_i} 1$ NAP, allowing 
their algorithm to solve such instances as well.

This algorithm
 relies upon the observation that the solution to 
$0 \xrightarrow{c_i} 1$ NAP for any clade can be obtained from the solutions
to its two child clades \cite{pardi07}.  Which solutions to use depends on
how the budget is allocated to the two subproblems.  If the budget
at $\mathcal{K}_e$ is $b$, 
then there are $b+1$ ways to split it across $\mathcal{K}_l$ and 
$\mathcal{K}_r$.  
By solving these $b+1$
pairs of subproblems, the optimal solution can be found in the pair
with maximum total $\mathbb{E}(PD)$ 
(plus the expected contribution of $e$).  Recursively proceeding in this 
fashion from the root down
would not yield an efficient algorithm as the number of possible
budget divisions increases exponentially with each level of the tree.
Instead, the clades are processed bottom-up from the leaves.  All $b+1$
scores are computed and stored in a dynamic programming table for each clade.
Each score can be determined by taking the maximum of $b+1$ possible 
scores of its child clades, which are already computed or computed directly
from (\ref{eq:epd}) if the clade contains a single leaf.  Each table entry can 
therefore be 
computed in $O(B)$ time.  There are $O(B)$ entries per clade and $O(n)$
clades in the tree giving a total running time of $O(nB^2)$.

\begin{figure}{
     \begin{center}\leavevmode
        \subfigure{%[Tree $T_1$]{
          %\label{X3CRedT1}
          \includegraphics[scale=0.35]{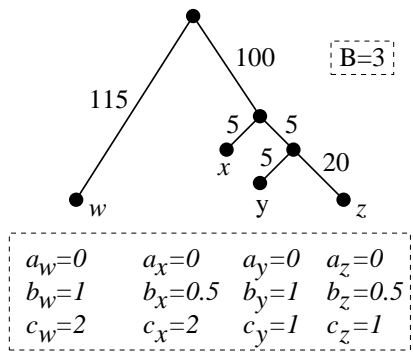}
        }
     \end{center}
  }
  \caption{An example why the dynamic programming algorithm of \cite{pardi07}
does not work for general instances of NAP.  The optimal allocation for the
clade containing $y$ and $z$ is not part of a globally optimal solution.}
  \label{fig:counter}
\end{figure}

This procedure 
does not work for $a_i \xrightarrow{c_i} b_i$
NAP because this version of the 
problem does not display the same optimal substructure \cite{pardi07}.  In 
$0 \xrightarrow{c_i} 1$, the dynamic programming algorithm implicitly
maximizes the survival probability of the clade in addition to its
$\mathbb{E}(PD)$ value.  The total score of the tree is a function of
both of these values which is why the algorithm works for this case.
In $a_i \xrightarrow{c_i} b_i$ NAP, a budget assignment that maximizes
survival probability of the clade does not guarantee that it will
have maximal $\mathbb{E}(PD)$ and vice versa. The correct allocation cannot
be made without knowledge of the entire tree; hence, the optimal
substructure exploited by \cite{pardi07} for $0 \xrightarrow{c_i} 1$ NAP
is not present. As an example, consider the instance of NAP in 
Figure~\ref{fig:counter} with $B=3$.  The optimal solution is to conserve
$w$ and $y$ for $\mathbb{E}(PD)=225$. However, locally computing the best 
allocation of budget $1$ for the clade containing $y$ and $z$ will select
$z$ for conservation, and any chance of obtaining the optimal solution will
be lost.  In this case, it is more important to maximize the 
survival probability
of the clade rather than $\mathbb{E}(PD)$, but there is no way for an
algorithm to be aware of this without globally solving the entire tree.  

\section{NAPX Algorithm}
\label{sec:ptas}

\subsection{Description}

In this section we present NAPX, an 
$O\left(\frac{nB^2h^2\left(\log{n} + \log{\frac{1}{\epsilon}}\right)^2}
{\log^2(1-\epsilon)}\right)$ 
dynamic programming algorithm for  $a_i \xrightarrow{c_i} b_i$
NAP that produces a ($1 - \epsilon$)-approximation of the optimal 
solution, where $h$ denotes the height of 
$\mathcal{T}$. As that in \cite{pardi07},
our algorithm is only polynomial if $B$ is polynomial in $n$.
  This assumption is justifiable if, for example,
$B$ is expressed in millions of dollars and its value will be a reasonably
small integer.  Without loss of generality, we also assume that no single
cost exceeds the budget.  

NAPX essentially generalizes the dynamic programming table of
\cite{pardi07} by computing for each clade, each desired survival probability
of the clade, and any budget between 1 and $B$, the maximum $\mathbb{E}(PD)$
score achievable while guaranteeing this survival probability.
  This way, we need not make the 
choice between maximizing $\mathbb{E}(PD)$ or probability as the tables
are constructed.  From the definition of
$\mathbb{E}(PD)$ in (\ref{eq:epd}), the probability of survival of
an edge can be written as a function of its two children.  Let $P_e$
denote the survival probability of edge $e$, and $l$ and $r$ be $e$'s children.
Then
\begin{equation}
\label{eq:pclade}
P_e = P_l + P_r - P_l P_r.
\end{equation}  
In the optimal solution for NAP on $\mathcal{T}$,
 assume $b$ dollars are assigned to clade $\mathcal{K}_e$ and 
$e$ survives with probability
$p$.  It follows that $i$ and $b-i$ dollars
are assigned to $\mathcal{K}_l$ and $\mathcal{K}_r$ respectively where
$0 \leq i \leq b$.  These subclades must survive with probabilities 
$j$ and $\frac{p-j}{1-j}$ (or 0 when $p=j=1$), 
for some $0 \leq j \leq p$, in order to 
satisfy (\ref{eq:pclade}). Because 
the probability is continuous, we discretize it into intervals by rounding it
down to the nearest
multiple of a chosen constant $\alpha$. Probabilities less than a chosen 
cutoff value $p_{min}$ are rounded to zero.
\begin{equation*}
  p \in \left\{ 0, \alpha^{\lceil \log_\alpha{p_{min}} \rceil}, ...,
  \alpha^2, \alpha,  1  \right\}
\end{equation*}
If two non-zero probabilities lie in the same interval, their ratio is at most 
$\alpha$. If they are in consecutive intervals, their ratio is likewise bounded
by $\alpha^2$.  For notational convenience, we define a mapping $\pi(\cdot)$
that rounds a probability to the lower bound of its corresponding interval.
\begin{equation*}
  \pi(p) =
  \begin{cases}
    0 & \text{if}~ p < p_{min} \cr
    \alpha^{\lceil \log_\alpha p \rceil} & \text{otherwise.}
  \end{cases}
\end{equation*}
We now formally describe our algorithm.  For each edge $e$, we construct
a two-dimensional table $T_e$ where $T_e(b,p)$ stores the optimal expected
diversity of $\mathcal{K}_e$ given that $b$ dollars are assigned to it and
 it survives with a probability that lies no less than $p$.
The table is constructed in the following manner if $e$ is a pendant edge
incident to the leaf for species $s$.
\begin{equation}
  \label{eq:dptable}
T_e(b,p) =  \begin{cases}
a_s \lambda(e) & \text{if $b < c_s$ and $p = \pi(a_s)$,} \cr
b_s \lambda(e) & \text{if $b \geq c_s$ and $p = \pi(b_s)$, or} \cr
-\infty & \text{otherwise.} \cr
\end{cases} 
\end{equation}
Otherwise, $T_e$ is computed from the tables of its two children, $T_l$ and 
$T_r$.
\begin{align}
\label{eq:dpinternal}
T_e(b,p) &= p \lambda(e) + \max_{i,j,k}\{ T_l(i,j) + T_r(b-i,k) \}
\end{align}
subject to 
\begin{align*}
i &\in \{0,1,2,...,b\}, \notag\\
j,k &\in  \{0,\alpha^{\lceil\log_\alpha\ p_{min} \rceil}, ...,
\alpha^2, \alpha,1\}, \notag\\ 
\pi(j + k-jk) &= p
\end{align*}

The $\mathbb{E}(PD)$ score
for the entire tree can be obtained by attaching an edge $e_r$ of length
0 to the root and finding $\max_j\{T_{e_r}(B,j)\}$.  
The tables are computed from the bottom up, and
each time an entry is filled, pointers are kept to the two entries 
in the child tables from which it was computed.  This way the optimal
budget allocation can be obtained by following the pointers down from 
the entry for the optimal score for $e_r$. 

\subsection{Approximation Ratio}
In this section, we express the worst-case approximation ratio as a function 
of the constants $p_{min}$ and $\alpha$ introduced above, 
beginning with $p_{min}$.  Note that 
since any species $s$ with $c_s > B$ can be 
transformed into a new species $s'$ with $c_{s'} = 0, b_{s'} = a_s$ and
$a_{s'} = a_s$ without
affecting the outcome, we can safely assume that $c_s \leq B$ for all 
$s \in X$. 
\begin{lemma}
  \label{lem:pmin}
Let $I$ be an instance of NAP for which there exists a constant $k$ such
that $b_i \geq n^{-k} \geq p_{min}$ for all $i \in S$. Consider 
a transformed instance $I'$ where
all $a_i$ values in the range $(0, p_{min})$ are rounded to 0. Let $OPT(I)$ 
and $OPT(I')$ be the expected PD scores of the optimal solutions to $I$ and
$I'$ respectively. Then
the ratio of these scores is bounded as follows: 
\begin{equation*}
OPT(I') \geq (1 - n^{k+1}p_{min})OPT(I)  
\end{equation*}
\end{lemma}
\begin{proof}
Let $\text{path}(s)$ be the set of edges comprising the path from leaf $s$ to
the root.  We define $w(s)$ as the expected diversity of the path from $s$
to the root if $s$ is conserved:
\begin{equation*}
  w(s) =  b_s \sum_{e \in \text{path}(s)}\lambda(e).
\end{equation*}
Let $w_{max} = \max_{s \in X}\{w(s)\}$.  
This value allows us to place a trivial lower bound on the optimal solution
(recalling that we can assume that $c_s \leq B$).
\begin{equation}
\label{eq:smaxbound}
  w_{max} \leq OPT(I).
\end{equation}
We also observe that if any species $s$ survives with a non-zero probability
smaller than $p_{min}$ in the optimal solution, its contribution to OPT($I$)
will be bounded by $\displaystyle\frac{p_{min} w(s)}{b_s}$.  It follows that 
\begin{equation*}
  OPT(I') \geq OPT(I) - \sum_{s \in X}\frac{p_{min} w(s)}{b_s}.
\end{equation*}
Since $b_s \geq n^{-k}$ and $w(s) \leq w_{max}$, we can 
express the bound as
\begin{equation*}
  OPT(I') \geq OPT(I) - n \frac{p_{min} w_{max}}{n^{-k}}.
\end{equation*} 
Dividing by OPT($I$) yields
\begin{equation*}
  \frac{OPT(I')}{OPT(I)} \geq 1 - \frac{n^{k+1} p_{min} w_{max}}{OPT(I)}.
\end{equation*}
From (\ref{eq:smaxbound}) we obtain
\begin{equation*}
  \frac{OPT(I')}{OPT(I)} \geq 1 - n^{k+1} p_{min}, 
\end{equation*}
which completes the proof.\qed
\end{proof}
The size of the probability intervals in the tables, determined by 
 $\alpha$, also affects the approximation ratio.  This relationship
is detailed in the following lemma.  
\begin{lemma}
\label{lem:rounding}
Let $OPT_e(b,p)$ denote
the optimal expected PD score for clade $\mathcal{K}_e$
if $e$ survives with 
probability exactly $p$ and $b$ dollars are allocated to it.
Now consider an instance of NAP such that all $a_s$ and $b_s$ are either 0 or 
at least $p_{min}$. For any $OPT_e(b,p)$ where $e$ is at height $h$ in
the tree, there exists a table entry $T_e(b,p')$
constructed by NAPX such that the following conditions hold:
\begin{eqnarray*}
i)& T_e(b,p') &\geq \alpha^hOPT_e(b,p) \\
ii)& p' &\geq \alpha^hp
\end{eqnarray*}
\end{lemma}
\begin{proof}
If $p=0$, then $OPT_e(b,p) = 0$ and the lemma holds.  For the remainder of the
proof, we assume $p \geq p_{min}$.  
The proof will proceed by induction on $h$, the height of $e$ in the tree,
beginning with the base case where $h=1$ and $e$ is a pendant connected to
leaf $s$.  We need only consider the cases where the optimal solution is
defined. So without loss of generality, assume we have $OPT_e(b,a_s) = 
\lambda(e)a_s$. From (\ref{eq:dptable}), we know there is an entry 
$T_e(b,\pi(a_s)) = a_s\lambda(e)$ and therefore both $i)$ and $ii)$ hold.

We now assume that the lemma holds for $h \leq x$ and consider some edge $e$
at height $x+1$. By definition, $OPT_e(b,p)$ can be expressed in terms of its
children $l$ and $r$.
\begin{equation*}
  OPT_e(b,p) = p \lambda(e) + OPT_l(i,j) + OPT_r(b-i,k)
\end{equation*}
where $j+k-jk=p$. From the induction hypothesis, there exist
\begin{align*}
  T_l(i,j') &\geq \alpha^xOPT_l(i,j) ~\text{and}\\
  T_r(b-i,k') &\geq \alpha^xOPT_r(b-i,k)
\end{align*}
where $j' \geq \alpha^xj$ and $k' \geq \alpha^xk$.  Let $q = j' + k' -j'k'$.
It follows that
\begin{equation}
  \label{eq:something}
q \geq \alpha^x j + \alpha^x k - \alpha^{2x}jk \geq \alpha^xp. 
\end{equation}
The left inequality in (\ref{eq:something}) 
holds because $j' + k' - j'k'$ increases
as $j'$ or $k'$ increase, so long as their values do not exceed 1. This
can be
 checked by observing that the partial derivatives with respect to $j'$ and
$k'$ are $1-k'$ and $1-j'$, respectively.

$T_l(i,j')$ and $T_r(b-i,k')$ will be considered when computing the entry
$T_e(b,p')$ where $p' = \pi(q)$. Since $q \geq p_{min}$, we have
 $\pi(q) \geq 
\alpha q$ because it simply rounds $q$ to the nearest multiple of $\alpha$. 
Therefore, $p' \geq \alpha^{x+1}p$ and $T_e(b,p')$ can  be expressed as follows.
\begin{align*}
  T_e(b,p') &\geq p' \lambda(e) +  T_l(i,j') + T_r(b-i,k')\\
  & \geq \alpha^{x+1}p \lambda(e) + \alpha^x OPT_l(i,j) + \alpha^x OPT(b-i,k)\\
  & \geq \alpha^{x+1}(p \lambda(e) + OPT_l(i,j) + OPT(b-i,k)\\
  & \geq \alpha^{x+1} OPT_e(b,p)
\end{align*}\qed
\end{proof}

Combining Lemmas~\ref{lem:pmin} and \ref{lem:rounding} allows us to state
that NAPX returns a solution that is at least a factor 
of $(1-n^{k+1}p_{min})\alpha^h$ of the optimal solution.  In this section we
show that these results also imply that a $(1-\epsilon)$ approximation can be
obtained in polynomial time for an arbitrary constant $\epsilon$.

\begin{lemma}
  \label{lem:tsize}
$\displaystyle  O\left(\frac{h \left(\log{n} + \log{\frac{1}{\epsilon}}\right)}
{|\log (1 - \epsilon)| }\right)$ 
probability intervals are required in the table in order to obtain 
a $1 - \epsilon$ approximation.
\end{lemma}
\begin{proof}
The number of probability intervals, $t$, required for the table is bounded by
the number of times 1 must be multiplied by $\alpha$ to reach $p_{min}$.  
Hence $\alpha^t \leq p_{min}$ and
\begin{equation}
  \label{eq:intervals1}
  t = \left\lceil  \frac{\log p_{min}}{\log \alpha} \right\rceil.
\end{equation}
From  Lemmas~\ref{lem:pmin} and \ref{lem:rounding} we can obtain the desired
approximation ratio by selecting $\alpha = \sqrt{(1-\epsilon)^{\frac{1}{h}}}$
and $p_{min} = \frac{1-\sqrt{1-\epsilon}}{n^{k+1}}$. Plugging these values into 
(\ref{eq:intervals1}) gives
\begin{align*}
t =  \left\lceil \frac{\log\left( \frac{1-\sqrt{1-\epsilon}}{n^{k+1}} \right)}
  {\log \left(\sqrt{(1-\epsilon)^{\frac{1}{h}}} \right)}\right\rceil =
\left\lceil \frac{2h(\log (1-\sqrt{1-\epsilon}) - (k+1)\log{n})}
  {\log (1-\epsilon)}\right\rceil
\end{align*}
It can be shown that $\log(1 - \sqrt{1-\epsilon})$ is $O(\log{\epsilon})$,  so 
multiplying by $\frac{-1}{-1}$ we can express $t$ asymptotically as 
\begin{equation*}
  t \in O\left(\frac{h \left(\log{n} - \log{\epsilon}\right)}
{-\log(1-\epsilon)}\right) 
=
O\left(\frac{h \left(\log{n} + \log{\frac{1}{\epsilon}}\right)}
{|\log(1-\epsilon)|}\right). 
\end{equation*}\qed
\end{proof}

\begin{theorem}
  \label{thm:ptas}
NAPX is a $(1-\epsilon)$-approximation with time complexity
\begin{equation*}
O\left(\frac{n B^2 h^2 \left(\log{n} + \log{\frac{1}{\epsilon}}\right)^2}
{\log^2(1 - \epsilon)}\right). 
\end{equation*}
\end{theorem}
\begin{proof}
For each table entry $T(b,p)$, we must find the maximum
of all possible combinations of
entries in the left and right child tables that satisfy $b$ and $p$.  These
combinations correspond to the possible $\{i,j,k\}$ triples from 
 (\ref{eq:dpinternal}).  There are $O(Bt^2)$ such combinations as $i$ 
corresponds to the budget and $j$ and $k$ correspond to probability intervals.
Furthermore, for fixed values of $p$ and $j$, there are potentially 
$O(t)$ different values of $k$ that could satisfy $\pi(j + k -jk)$ due to
rounding. It follows that a naive algorithm would have to compare all
$O(Bt^2)$ combinations when computing the maximum in (\ref{eq:dpinternal})
for each table entry. 

Fortunately, because $\pi(j + k -jk)$ is 
monotonically nondecreasing with respect to either
$j$ or $k$, we can directly compute for any fixed $p$ and $j$ the
interval of $k$ entries that satisfy $\pi(j + k - jk) = p$:
\begin{equation*}
  \left[ \left\lceil \log_\alpha \left( \frac{\alpha p - j}{1 - j} \right) 
      \right\rceil,
    \left\lceil \log_\alpha \left( \frac{p - j}{1- j} \right) \right\rceil
    \right).
\end{equation*}
Finding the value of $k$ in the interval such that $T(b-i, k)$ is maximized
is effectively a range maxima query (RMQ) on an array. Regardless of the size 
of the interval, such a query can be performed in constant time if instead
of an array, the values are stored in a RMQ structure as described in 
\cite{bender00}.  Such structures are linear both in space and the time they
take to create, meaning that we can use them to store each column in the table 
(corresponding to budget value $i$) without adversely affecting the complexity.
Now, when given a pair $\{i,j\}$, the optimal value of $k$ can be computed in
constant time, bringing the complexity of filling a single table entry to
$O(Bt)$, the number of combinations of the pair $\{i,j\}$.   

There are $O(Bt)$ entries in each table, and a table for each of the $O(n)$
edges in the tree. The space complexity is therefore $O(nBt)$ and the time
complexity is $O(nB^2t^2)$. Substituting $t$ for the value that yields
a $(1 - \epsilon)$ approximation ratio shown in
Lemma~\ref{lem:tsize} gives
$\displaystyle O\left(\frac{nB^2h^2\left(\log{n} + \log{\frac{1}{\epsilon}}\right)^2}
{\log^2(1-\epsilon)}\right)$.

\end{proof}

\subsection{Expected Running Time}

Since in general
 the height of a phylogenetic tree with $n$ leaves is $O(n)$, the running
time derived above is technically cubic in $n$.  Fortunately, for most 
inputs we can expect the height to be much smaller.  In this section, we
will provide improved running times for trees generated by the two principal
random models.  Additionally we will show that caterpillar trees, which
should be the pathological worst-case topology according to the 
above analysis, actually have a much lower complexity.

The Yule-Harding model \cite{yule25}\cite{harding71}, also known as the 
equal-rates-Markov model, assumes that trees are formed by a succession
of random speciation events.  The expected height of trees formed in this
way, regardless of the speciation rate, is $O(\log{n})$ \cite{erdos99} giving
a time complexity of $O\left(\frac{nB^2\log^2n\left(\log{n} + \log{\frac{1}{\epsilon}}\right)^2}
{\log^2(1-\epsilon)}\right)$.

A caterpillar tree is a tree where all internal nodes are on a path beginning
at the root, and therefore has height $n$. This implies 
that every internal edge has at least one child edge that is incident to a 
leaf.  Suppose edge $e$ has child $l$ that is incident to the leaf for 
species $s$.  This 
table only contains two meaningful values: $T_l(0, a_s)$ and $T_l(c_s, b_s)$.
Therefore to compute entry $T_e(b,p)$, only $O(1)$ combinations of child table
entries need to be compared and the time complexity is improved to
$O\left(\frac{n^2B\left(\log{n} + \log{\frac{1}{\epsilon}}\right)}
{|\log(1-\epsilon)|}\right)$.

\section{Conclusion}

NAPX is, to our best knowledge, the first polynomial-time algorithm for 
$a_i\xrightarrow{c_i}b_i$ NAP that places guarantees on the approximation
ratio. While there are still some limitations, especially for
 large budgets or tree heights, our algorithm still significantly 
increases the
number of instances of NAP that can be solved.  Moreover, our expected
running time analysis shows that the algorithm will usually be much more
efficient than its worst-case complexity suggests.  This
work towards a more general solution is important if the Noah's Ark Problem
 framework
is to be used for real conservation projects.  Some interesting questions do
remain, however.  Does NAP remain NP-Hard when the budget is constrained to
be polynomial in $n$?  We conjecture that it is, but the usual reduction from
Knapsack is clearly no longer valid. We would also like to find
an efficient algorithm whose complexity is independent of $h$ and/or $B$.

\bibliography{napx_wabi08}

\end{document}